# FLIGHT TRAJECTORY RECREATION AND PLAYBACK SYSTEM OF AERIAL MISSION BASED ON OSSIMPLANET

Wu Wu[1] , Jiulin Hu[2], Xiaofang Huang[3], Huijie Chen[4],Bo Sun[5]

College of Information Science and Technology, Beijing Normal University,Beijing, China

tosunbo@bnu.edu.cn

## ABSTRACT

*Recreation of flight trajectory is important among research areas. The design of a flight trajectory recreation and playback system is presented in this paper. Rather than transferring the flight data to diagram, graph and table, flight data is visualized on the 3D global of ossimPlanet. ossimPlanet is an open-source 3D global geo-spatial viewer and the system realization is based on analysis it. Users are allowed to choose their interested flight of aerial mission. The aerial photographs and corresponding configuration files in which flight data is included would be read in. And the flight statuses would be stored. The flight trajectory is then recreated. Users can view the photographs and flight trajectory marks on the correct positions of 3D global. The scene along flight trajectory is also simulated at the plane's eye point. This paper provides a more intuitive way for recreation of flight trajectory. The cost is decreased remarkably and security is ensured by secondary development on open-source platform.*

## 1. INTRODUCTION

Flight trajectory is important in the flight collision, flight planning, flight accidents investigation and flight simulation areas. Generally, the recreation of flight trajectory [1-2] is to transfer the flight data to the diagram, graph, table, etc. It is important in aircraft safety assessments, aircraft maintenance, accident investigation and event analysis. However, the results by these common solutions are not intuitive for the users [3]. The flight data, e.g., the flight angles, positions and m Accordingly, it is a better way to recreate the flight trajectory by visualizing the flight data.

The visual simulation technology makes it possible. Flight Viz [4] produced by SimAuthor Company could playback the flight trajectory in 3D visual effects, but its cost is very high. The 3D Flight Simulation System EasyFlight developed by China Academy of Civil Aviation Science and Technology is a professional aeronautic platform [5]. It's mainly used in flight simulation of major accidents, but rarely in common flight trajectory recreation. Yong Tang [6] presented a solution for 3D flight trajectory and 6-DOF flight simulation based on Google Earth [7]. But the research result relies much on the server of Google Earth, thus the users may concern about the security and the cost.

From the point of view of the cost and security, secondary development on open-source platform is a better choice. In this paper, ossimPlanet [8] is taken as the development platform. It is an accurate 3D global geo-spatial viewer that is built on the OSSIM [9], OpenSceneGraph [10], and Trolltech QT[11] open source software libraries [12]. It could provide accurate 3D global visualization and collaboration [13], and has the following three advantages: (1) It's open-source. It costs less than platform, e.g., Google Earth. Especially, we can realize more customized functions and ensure its security. (2) It's built on OSSIM, which has a powerful suite of geospatial libraries and applications to process imagery, maps, terrain, and vector data [9]. This paper focuses on the flight trajectory of aerial mission, which includes a lot of aerial photographs. Thus, OSSIM can provide strong support on image processing. (3) any other flight parameters could describe the spatial status of flight.

It's written in C++ and thus has higher performance than the platforms written in other languages, such as World Wind written in C#.

In this paper, a flight trajectory recreation and playback system of aerial mission will be implemented based on ossimPlanet. The system would recreate and playback the trajectory on 3D global, thus it will be more intuitive. The development on ossimPlanet ensures the security in a low cost and high performance. This paper is organized as follows. In Section II, the requirement analysis of the whole system is presented. The key problems and the corresponding solutions are given in Section III. The system realization is introduced in Section IV. The simulation results are shown in Section V. The conclusions are summarized in Section VI.

## 2. REQUIREMENT ANALYSIS

The following formatting rules must be followed strictly. This (.doc) document may be used as a template for papers prepared using Microsoft Word. Papers not conforming to these requirements may not be published in the conference proceedings.

The system functions are shown in Fig.1, and their detailed descriptions are given below.

- Choose flight. Users are allowed to choose their interested flight, that is, to fix the local path where the aerial photographs and corresponding configuration files are located. Then these files would be read in, and the statuses of the plane are stored after the necessary processing on these input files.

- Observe photographs. Users are allowed to observe the input photographs. These photographs would be pasted on their correct positions which are set in the configuration files.

- Observe trajectory. Users are allowed to observe the flight trajectory on the 3D global of ossimPlanet. Both the input flight trajectory points and the interpolated trajectory points are marked on the 3D global.

- Observe simulation. Users could follow the plane's eye point to view the flight trajectory dynamically.

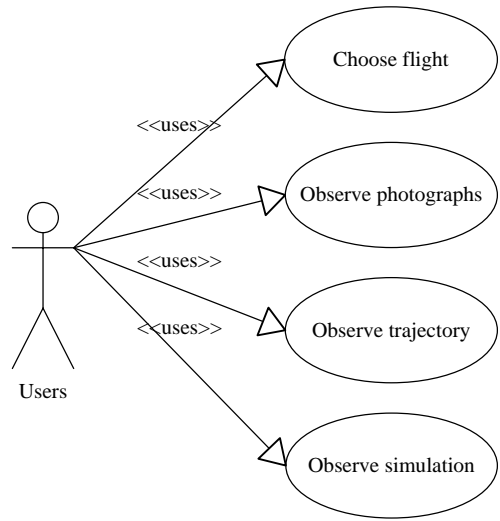

Fig. 1.   Use case diagram.

## 3. KEY PROBLEMS

There are three key problems for the system realization. (1) Data processing. The input data includes the aerial photographs and configuration files. Every aerial photograph has a corresponding configuration file in which the flight data is included, e.g., the flight statuses, flight positions, pilot's operations, etc. The required flight data will be taken for interpolation. (2) Data display. It is to display the aerial photographs and mark flight trajectory points on the 3D global of ossimPlanet. (3) Flight trajectory playback. It is to playback the flight trajectory on the 3D global of ossimPlanet.

### 3.1. Data Processing

Without loss of generality, we make following assumptions on the motion of plane:  (1) It's rigid body motion. (2) The translation is with the centroid and the rotation is around the centroid.

To describe the motion clearly, we should take proper flight data from the configuration files. Generally, 6 degree-of–freedom (DOF), i.e., 3 position coordinates (longitude, latitude and height) and 3 posture angles (heading angle, pitch angle and roll angle), is always used [6]. The posture angles are shown in Fig.2.

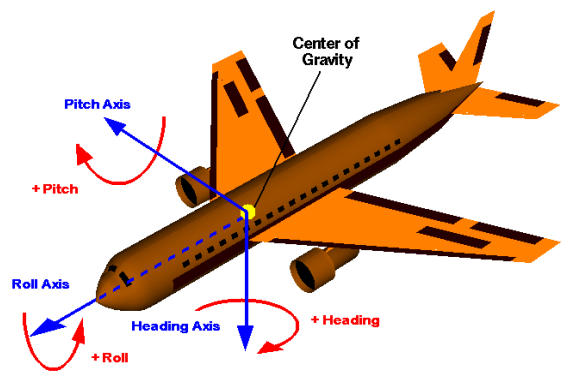

Fig. 2.   The heading, roll and pitch angle of plane [14].

After we get the flight trajectory points contained the 6 DOF parameters, it's necessary to smooth the flight trajectory by interpolation. The interpolation on the position coordinates and posture angles will be done respectively. Since it's shown that the unstable heights may result in flight collision [15], the height is assumed invariable in position coordinates and not interpolated. De Boor's algorithm [16] will be used to interpolate the longitude and latitude. In the posture angles, the heading angle is changed with the stress of plane [17] and always very small. The changes of pitch angle in real aerial mission are generally less than 5 degrees and roll angle is always 0 degree [18]. In this paper, we use linear interpolation for the posture angles smoothing.

## 3.2. Data Display

### 3.2.1. Photograph Display

To display the photographs on the 3D global of ossimPlanet, the corresponding geometry files for photographs are required. A geometry file instance is given in Fig. 3.In which the projection type, datum, longitude, latitude and some other geographic parameters are set.

```
type: ossimEquDistCylProjection
origin_latitude:0.0
central_meridian: 0.0
pixel_scale_units: degrees
pixel_scale_xy: ( .133, .133 )
datum: WGE
tie_point_units: degrees
tie_point_xy: (-180.0, 90.0)
pixel_type: area
```

Fig. 3.    Geometry file instance.

In Fig. 3, the type defines the projection of the photograph, and the default projection of ossimPlanet is cylindrical equidistant projection [19]. The origin_latitude and the central_meridian are always 0 degree. The pixel_scale_xy is the actual scale of every pixel of the photograph and its unit is defined in the pixel_scale_units. The tie_point_xy is the coordinate of the photograph as (longitude, latitude) and its unit is defined in tie_point_units.

After the photographs and configuration files are read in, the corresponding geometry files are created according to the configuration files. Then by using ossimPlanet's API, the photographs could be displayed on the 3D global.

### 3.2.2. Flight Trajectory Display

Keyhole Markup Language (KML) [20] is a Markup Language to describe and store geographical information, such as point, line, surface, three-dimensional models, etc. A KML file instance is given in Fig.4.

Generally, a KML file includes 3 parts: (1) XML Header; (2) The definition of KML namespace; (3) The object of geographical indication [21]. In Fig.4, <Stlye> indicates a style may be used for objects and <Placemark> indicates a place mark. The KML file in Fig.4 indicates a point at (121.48844, 53.332649, 0), and it will be shown as an icon whose hyperlink is given in the referenced link.

```
<?xml version="1.0" encoding="UTF-8"?>
<kml xmlns="http://www.opengis.net/kml/2.2"
    xmlns:gx="http://www.google.com/kml/ext/2.2">
 <Document>
<Style id ="style60" >
 <IconStyle>
 <Icon><href>reference link</href></Icon>
 </IconStyle>
 </Style>
  <Placemark>
    <styleUrl>#style60</styleUrl>
    <Point>
     <coordinates>121.48844,53.332649,0</coordinates>
     </Point>
    </Placemark>
 </Document>
</kml>
```

Fig. 4.   KML file example.

A KML file is created for the input trajectory points and interpolated trajectory points. Then by using ossimPlanet's API, the KML file could be loaded and thus the trajectory points could be marked on the 3D global.

### 3.3. Flight Trajectory Playback

To playback the flight trajectory, it's necessary to have the knowledge of the 3D world of ossimPlanet. (1) Coordinate Systems and Transformations. In the 3D world, the basic work is to confirm the coordinate systems and find out the coordinate transformations. (2) View Transformation. To playback flight trajectory is to change the eye point with the flight status. Thus the view transformation is important. (3) Rendering theory of ossimPlanet. The actual development work should be based on the rendering theory of ossimPlanet.

### 3.3.1. Coordinate Systems and Transformations

The Geographical Coordinate System, World Coordinate System and Local Coordinate System are briefly introduced as follows:

a)   Geographical Coordinate System. In this coordinate system, each point is determined by its longitude, latitude and the height above a WGS-84 reference ellipsoid [22].

b)   World Coordinate System. The world coordinate of ossimPlanet is Earth Centered Earth Fixed (ECEF) [23], as the XYZ coordinate system shown in Fig.5.

c)   Local Coordinate System. The local space reference (LSR) system of ossimPlanet is as UVW coordinate system shown in Fig.5.

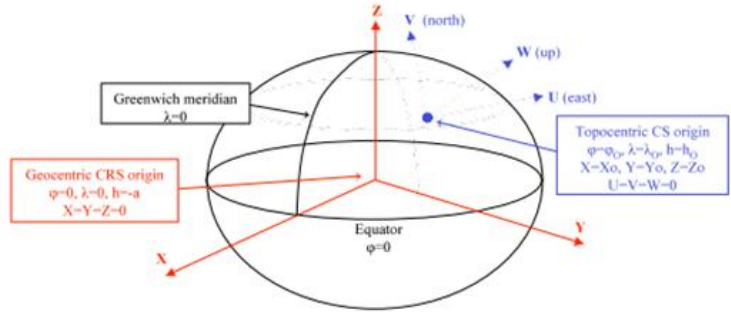

Fig. 5. Coordinate systems of ossimPlanet[24].( λ is longitude ,φ is latitude and h is the height )

Both the Geographical Coordinate and the Local Coordinate could be converted into the World Coordinate as follows [24].

a) From Geographical Coordinate to ECEF

$$X = (\nu + h)\cos\varphi\cos\lambda$$

$$Y = (\nu + h)\cos\varphi\sin\lambda \qquad\qquad (1)$$

$$Z = [(1 - e^2)\nu + h]\sin\varphi$$

b) From LSR to ECEF

$$X = X_0 - U\sin\lambda_0 - V\sin\varphi_0\cos\lambda_0 + W\cos\varphi_0\cos\lambda_0$$

$$Y = Y_0 + U\cos\lambda_0 - V\sin\varphi_0\sin\lambda_0 + W\cos\varphi_0\sin\lambda_0 \qquad (2)$$

$$Z = Z_0 + V\cos\varphi_0 + W\sin\varphi_0 ,$$

where ν is normal vector of latitude φ and its value is $\nu = a/(1 - e^2\sin^2\varphi)^{0.5}$, h is the height above the surface of ellipsoid, e is eccentricity and $e^2 = (a^2 - b^2)/a^2 = 2f - f^2$, φ is latitude and λ is longitude, a is semi-major axis, b is semi-short axis and f is flattening.

### 3.3.2. View transformation

From the general process of 3D graphics display [25], we can easily convert the World Coordinate to the View Coordinate as follows,

$$ViewCoord = WorldCoord * \mathrm{VM} * \mathrm{PM} * \mathrm{WM} , \qquad (3)$$

where VM is the view matrix, PM is the projection matrix and WM is the window matrix.

In ossimPlanet, PM and WM in (3) are fixed. Therefore, VM should be calculated for the view transformation. That is to place the eye point of ossimPlanet on the proper position in proper posture. The position of eye point is determined by the position of the plane given in the configuration files. Then we can convert the position coordinate in the Geographic Coordinate System $GeoEye(l,l,h)$ to that in the World Coordinate System $WorldEye(x_0, y_0, z_0)$ as in (1).

From the posture angle, we can get the rotation matrix of the eye point:

$$RotateMatrix = R_z(h) * R_y(p) * R_x(r),  \qquad (4)$$

where h is heading angle, p is pitch angle and r is roll angle. $R_z(h)$, $R_y(p)$ and $R_x(r)$ are the corresponding rotation matrixes [25].

Since RotateMatrix is in the Local Coordinate System, we should convert it to the World Coordinate System as in (2) and have the rotation matrix of the eye point in LSR,

$$RotationLsrMatrix = RotateMatrix * LsrMatrix, \qquad (5)$$

then, VM is as follows:

$$ViewMatrix = RotationLsrMatrix * WorldEye, \qquad (6)$$

### 3.3.3. Rendering Theory of ossimPlanet

The rendering circle [26] of ossimPlanet is to loop the frame() before the scene is finished. Every frame has the following three traversals:

a) Event Traversal. This part is implemented in eventTraversal(), where the different kinds of events are handled. The events include the mouse events, keyboard events, windows, callbacks of cameras, etc.

b) Updating Traversal. This part is implemented in updateTraversal(), where the updating callbacks are traversed and executed .

c) Rendering Traversal. This part is implemented in renderingTraversal(), where the rendering work such as the Cull and the Draw are done.

The basic rendering progress in ossimPlanet is described below. (1) The events from GUI or the scene are caught and handled in Event Traversal. The corresponding scene parameters are also calculated. (2)The scene parameters calculated by the Event Traversal are updated in Updating Traversal. (3)The scene parameters updated in Updating Traversal would be shown by Rendering Traversal. Thus, the scene would change with the events through the cooperation of the 3 traversals.

In ossimPlanet, the rendering circle is finished in a component called Viewer. It's shown in Fig. 6 that the Viewer includes Manipulator, GUI Event Handler, Scene and Camera. The Manipulator is an instance for roaming in the scene of the Viewer. All of the events from the GUI or the scene will be collected in the Manipulator. And these event messages will be translated to and finally handled in Navigator. In Navigator, the calculations of scene parameters of Event Traversal are completed.

To implement playback of flight trajectory in ossimPlanet, we should add our own event handlers in Event Traversal. And the handlers would handle customized events and calculate the scene parameters as we design. The rest work could be then done by the other 2 traversals.

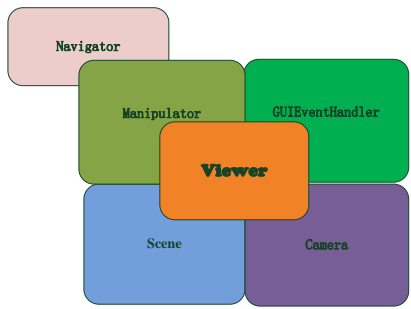

Fig. 6.   Viewer of ossimPlanet.

# 4. SYSTEM REALIZATION

Based on the analysis in Section III, the system realization is mainly to overwrite the Manipulator and Navigator of ossimPlanet. The user interactions are through the GUI (ossimPlanetQtMainWindow) and the corresponding event handlers are added in Navigator following the theory of Event Traversal. To store the information of flight trajectory, data structures are designed.

## 4.1. Component Design

The component diagram of the whole system is shown in Fig.7 and the corresponding description is summarized below.

GUI (ossimPlanetQtMainWindow) provides 3 interfaces for user interactions. on_viewStartInputtingPath_triggered() is for inputting the interested flight of aerial mission, on_flieOpenKml_triggered() is for displaying the flight trajectory and on_viewShowThePath_triggered() is for simulating the flight trajectory playback. After users' operations, GUI will translate event messages to the Navigator which is bridged by the Manipulator. Then, Navigator will handle these events.

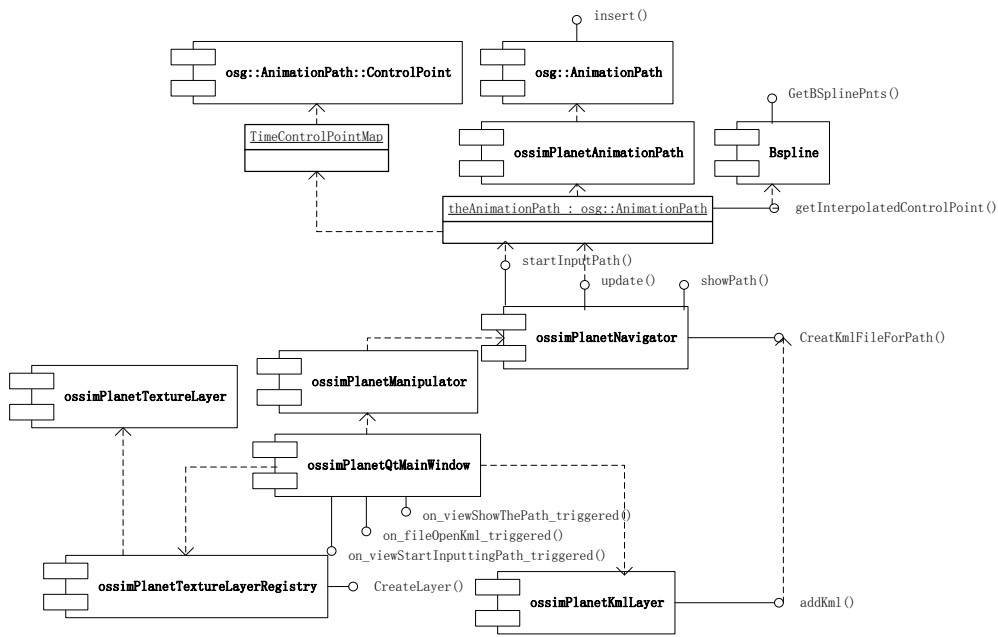

Fig. 7.   Component diagram.

Interfaces (startInputPath(), CreateKmlFileForPath() and showPath()) are added to the Navigator. They are the responses to the GUI events. The startInputPath() is to receive the raw files (photographs and configuration files) and store them. The CreateKmlFileForPath() is to create a KML file and write the points of flight trajectory into it. The showPath() is to change the rendering mode of ossimPlanet to simulation mode. The calculation of scene parameters is done in the update().The ossimPlanetTextureLayer is the layer of photographs and the ossimPlanetKmlLayer is the layer of the KML files. Every photograph or KML file shown on the 3D global of ossimPlanet corresponds to a layer.

## 4.2. Data Structures

The data structures used for storing the flight trajectory points are shown in Fig.8, and the corresponding description is given below. (1) Struct PathPoint. It includes 6 attributes corresponding to the 6 DOF parameters and denotes the input flight trajectory point. An array of the PathPoint denotes the flight trajectory points gotten from the configuration files. (2) Class BSpline. It is for the de Boor's interpolation algorithm and shown in Fig.8 (a). (3) Class ControlPoint. It denotes the flight trajectory point after interpolation. And it has 3 attributes, where the _position is the position of eye point in the world coordinate system, the _rotation is the rotation of the eye point and the _scale is scale factor. The view matrix could be set by them. It's shown in Fig.8 (b). (4) Class ossimPlanetAnimationPath. It is to store the flight trajectory points after interpolation and is shown in Fig.8 (c). In which map<double,ControPoint> is to store the mapping relationship between the flight trajectory point and its relative time.

| BSpline |
| --- |
| #ShapePoints |
| #NodeVector |
| #MyControlPoints |
| #BSplinePoints |
| +BSpline() |
| +CalNodeVector() |
| +CalControlPnts() |
| +GetdeBoorValue() |
| +CalBSplinePnts() |
| +GetBSplinePnts() |

| ControlPoint |
| --- |
| #osg::Vec3d _position |
| -osg::Quat _rotation |
| -osg::Vec3d _scale |
| +ControlPoint() |
| +setPosition() : void |
| +getPosition() |
| +setRotation() |
| +getRotation() |
| +setScale() |
| +getScale() |
| +getMatrix() |

Fig.8 (a). Class BSpline.          Fig.8 (b).Class ControlPoint.

| ossimPlanetAnimationPath |
| --- |
| +ControlPoint |
| +map<double,ControlPoint> TimeControlPointMap |
| # TimeControlPointMap _timeControlPointMap |
| +getMatrix() : bool |
| +getInterpolatedControlPoint() : bool |
| +insert() : void |
| +getFirstTime() : double |
| +getLastTime() : double |
| +getPeriod() : double |
| +getTimeControlPointMap() |

Fig.8 (c). Class ossimPlanetAnimationPath.

Fig. 8.  Main data structures.

### 4.3. Storage of Interested Flight

After users input the interested flight from GUI, the on_viewStartInputtingPath_triggered() is triggered. The cooperation diagram is shown in Fig.9.

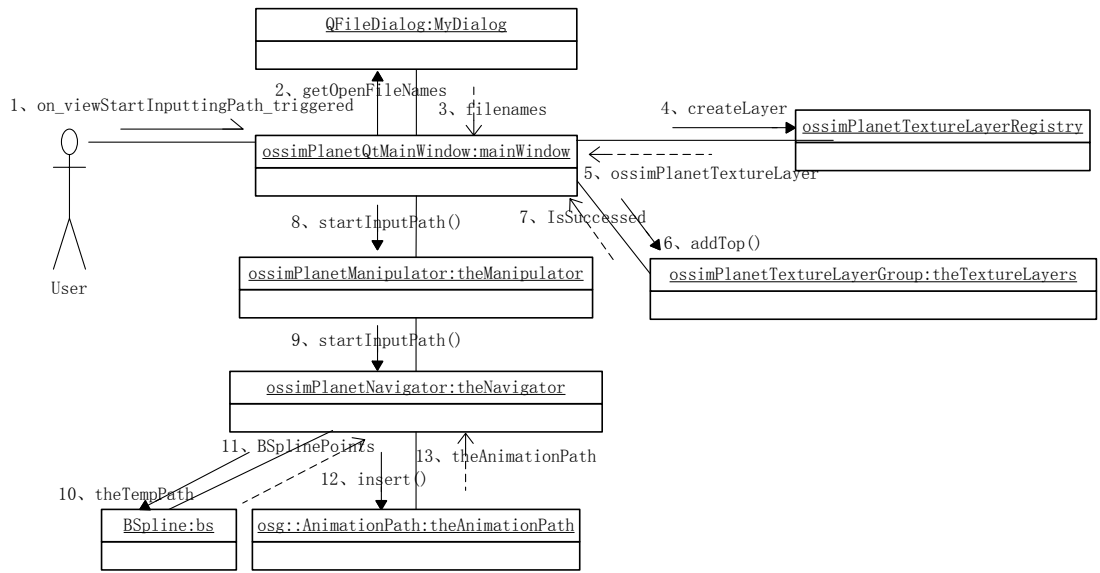

Fig. 9. Cooperation of the components after inputting interested flight.

The message translation and cooperation are summarized below.

- Users fix the path of the aerial photographs and configuration files. Then the point of flight trajectory is stored as PathPoint and the input trajectory points as an array of PathPoint. The geometry files are also created.

- createLayer() is invoked, and the texture layers corresponding to the photographs are created.

- addTop() is invoked to load these layers. The actual locations of the photographs are defined in the corresponding geometry files so that the photographs are displayed in the correct positions.

- GUI translates the event messages to the Navigator through the Manipulator.

- In the Navigator, the input flight trajectory points are interpolated and the results are stored as ossimPlanetAnimationPath.

### 4.4. Display of Flight Trajectory

After users choose to display the flight trajectory from GUI, the on_flieOpenKml_triggered() is triggered. The cooperation diagram is shown in Fig. 10, and the message translation and cooperation are summarized below.

- The users' option is translated to the Navigator through the Manipulator.

- In the Navigator, the CreateKmlFilePath() is invoked to create a KML file.

- Traverse the input trajectory points and export them into the KML file.

- Traverse the interpolated trajectory points and export them into the KML file.

- The addkml() in ossimPlanetKmlLayer is invoked to create a corresponding KML layer on ossimPlanet for display.

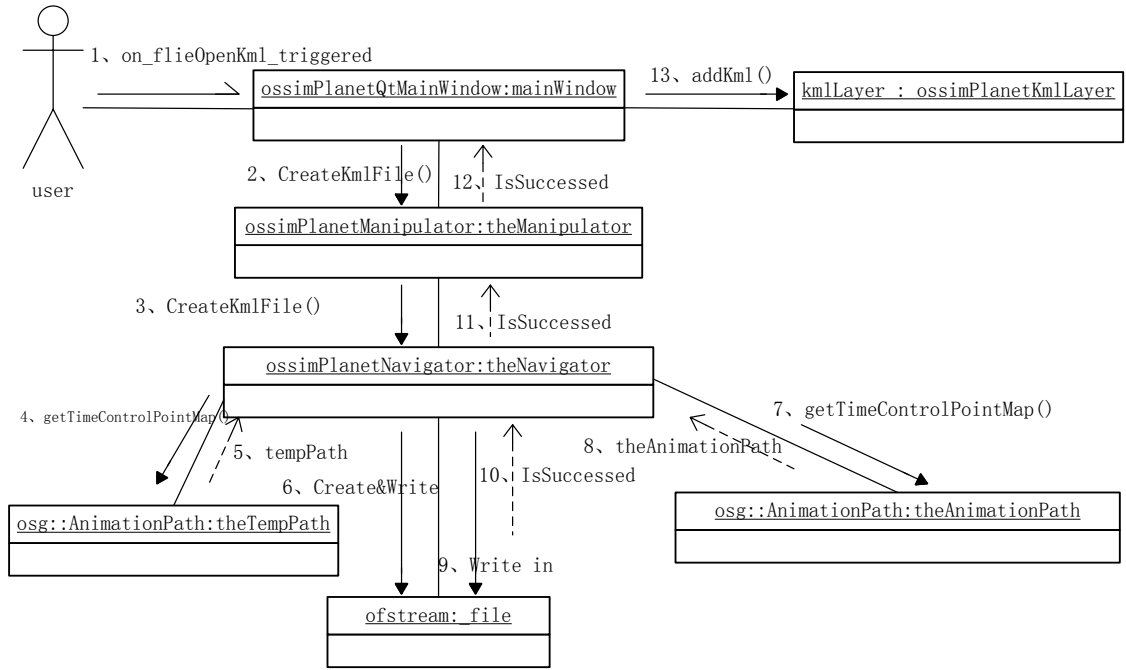

Fig. 10. Cooperation of the components after choosing to display trajectory.

## 4.5. Playback of Flight trajectory

After users choose to playback the flight trajectory, on_viewShowThePath_triggered() is triggered. The cooperation diagram is shown in Fig.10. The message translation and cooperation are summarized blow.

- The users' option is translated to the Navigator through the Manipulator.

- In the Navigator, the showPath() is invoked. Current rendering mode is set as the simulation mode.

- The rest work is done along with the Event Traversal. In every frame, the scene parameters are calculated in update(). And the calculation is done according to the ControlPoint in the map of ossimPlanetAnimationPath .

- Loop the third step with the refresh of ossimPlanet to playback the flight trajectory dynamically.

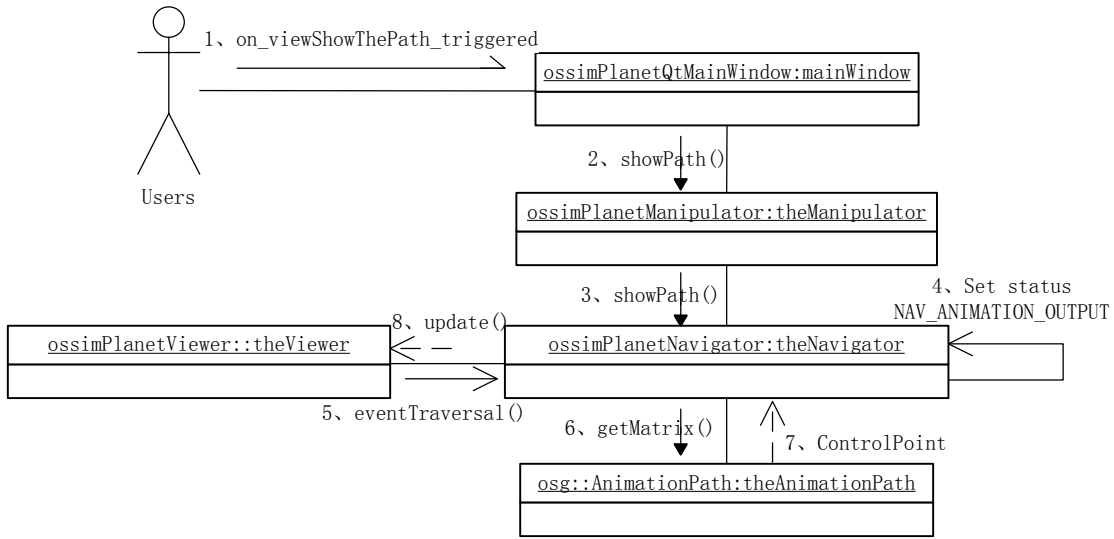

Fig. 11. Cooperation of thecomponents after choosing to playback the flight trajectory.

# 5. SIMULATION RESULTS

The simulation is done on the ossimPlanet 1.8.4. The interested flight data includes ten photographs and their configuration files. The 10 red-dotted points are the input flight trajectory points as shown in Fig.12 and the corresponding photograph is pasted as shown in Fig.13. The green marks are the interpolated flight trajectory points.

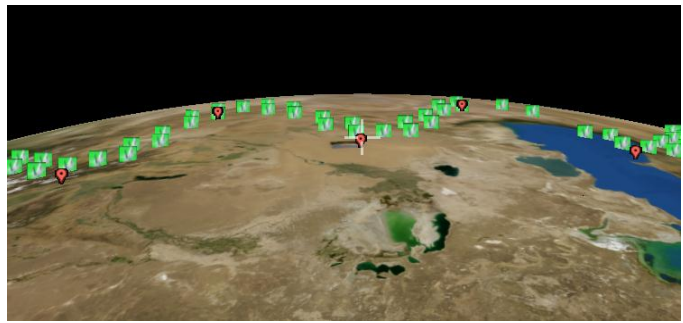

Fig. 12. Marks of flight trajectory.

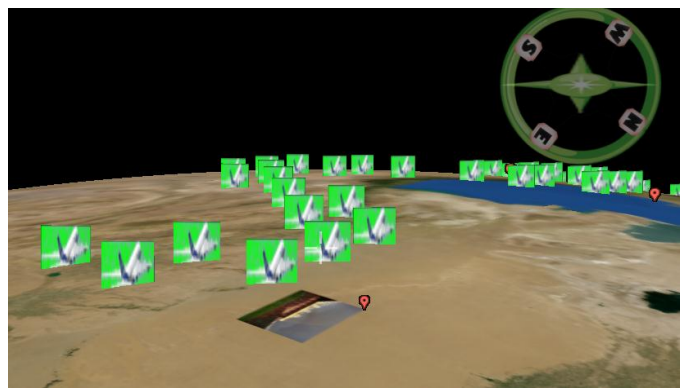

Fig. 13. Playback of flight trajectory.

During playback of flight trajectory, the eye point changes along the flight trajectory. As shown in Fig.13, users could view the photographs and the trajectory marks on the 3D global dynamically. The eye point will change with the plane when the ossimPlanet refreshes its scene.

# 6. CONCLUSIONS

In this paper, flight trajectory recreation and playback system of aerial mission is implemented based on open-source 3D global platform – ossimPlanet. Users can choose their interested flight of aerial mission. Then the aerial photographs would be displayed on the proper geographic positions of ossimPlanet. The flight trajectory also would be recreated and marked. In addition, the playback of the flight trajectory is simulated on ossimPlanet. These functions allow users to analyze their interested flight in a more institutive way.

The development on open-source platform ensures the security of system in a low cost and high performance. Especially, it allows developers to implement more customized functions. During the development, APIs of ossimPlanet about loading images, loading KML files and rendering frame are overwritten. This paper provides a general method for the development on ossimPlanet with its rendering theory.

**Authors**

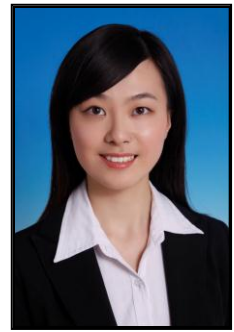

Wu Wu is a graduate student for master at the Research Center for Natural Computing and Software, College of Computer Science and Technology, Beijing Normal University. Her research interests include software engineering and information systems,3D GIS.Her email is midou@mail.bnu.edu.cn;